\documentclass[5p]{elsarticle}

 \usepackage{graphicx}
 \usepackage[font=footnotesize]{subfig} 

\usepackage{amssymb}
\usepackage{color}

\newcommand{\Omit}[1]{}

\def\beq{\begin{equation}}
\def\eeq{\end{equation}}
\def\bea{\begin{eqnarray}}
\def\eea{\end{eqnarray}}
\def\vd{\langle v_d q \rangle}

\newcommand*{\eqref}[1]{Eq.~(\ref{eq:#1})}
\newcommand*{\eqlab}[1]{\label{eq:#1}}
\newcommand*{\figref}[1]{Fig.~\ref{fig:#1}}
\newcommand*{\figlab}[1]{\label{fig:#1}}

\newcommand*{\secref}[1]{Section~\ref{sec:#1}}
\newcommand*{\seclab}[1]{\label{sec:#1}}

\def\vyp#1#2#3{{\bf #1}, #3 (#2)}  

\def\pl#1#2#3{Phys.~Lett.~\vyp{#1}{#2}{#3}}

\begin{document}

\begin{frontmatter}



\title{The Lateral Distribution Function of Coherent Radio Emission from Extensive Air Showers;
Determining the Chemical Composition of Cosmic Rays}


\author[KVI]{Krijn D. de Vries}
\author[KVI]{Ad M. van den Berg}
\author[KVI]{Olaf Scholten}
\ead{scholten@kvi.nl}
\author[SUBA]{and Klaus Werner}
\address[KVI]{Kernfysisch Versneller Instituut, University
of Groningen, 9747 AA, Groningen, The Netherlands}
\address[SUBA]{SUBATECH,
University of Nantes -- IN2P3/CNRS-- EMN,  Nantes, France}

\begin{abstract}
The lateral distribution function (LDF) for coherent electromagnetic radiation from air
showers initiated by ultra-high-energy cosmic rays is calculated
using a macroscopic description. A new expression is derived to calculate the coherent radio pulse at
small distances from the observer. It is shown that for small distances to the shower axis
the shape of the electric pulse is determined by the `pancake' function, describing the
longitudinal distribution of charged particles within the shower front, while for large
distances the pulse is determined by the shower profile. This reflects in a different
scaling of the LDF at small and at large distances. As a first application we calculate the
LDF for proton- and iron-induced showers and we show that this offers a very sensitive measure
to discriminate between these two. We show that due to interference between the geo-magnetic and the
charge-excess contributions the intensity pattern of the radiation is not circular symmetric.
\end{abstract}

\end{frontmatter}


\section{Introduction}

In recent years much progress has been made modeling electric pulses initiated by extensive air
showers (EAS)~\cite{Hue08,Sch08,Wer08}. One of the reasons for this progress are the results from
the LOPES~\cite{Fal05,Ape06} and CODALEMA~\cite{Ard06,Ard09} experiments. Both experiments
indicate that the dominant emission mechanism is due to induction effects from the Earth's
magnetic field which exerts a Lorentz force on the charged particles in the shower. The
emission process can be described in a microscopic model where the individual electrons and
positrons move on cyclotron orbits~\cite{Hue08,Fal03,Sup03,Hue05}. The importance of
coherent radio emission was already noted in earlier research on radio emission from air
showers~\cite{Jel65,Por65,Kah66,All71}. Recently, this approach has received renewed
attention with the realistic calculations in the Macroscopic Geo-Magnetic Radiation (MGMR)
model as presented in Ref.~\cite{Sch08}.

The experimental results have triggered plans for an extensive array of radio detectors at
the Pierre Auger Observatory~\cite{Ber07,Cop09,Rev09}. A clear theoretical understanding of the pulse
shape and its dependence on the distance from the shower axis is therefore of importance.
In the present paper we derive a new expression for the electromagnetic pulse that is more
appropriate for observer positions that are closer to the shower axis while the expression derived in
Ref.~\cite{Sch08} is appropriate for large distances. This allows us to address
quantitatively the differences in the structure of the electromagnetic pulse emitted by the
EAS at large and small distances. It was already noticed in an
earlier study~\cite{Sch08} that in a macroscopic calculation the structure of the
electromagnetic pulse at small distances is strongly affected by the distribution of the
particles in the shower front (the pancake). In Ref.~\cite{Arena} the importance of length
scales was emphasized in the understanding of macroscopic geo-magnetic radiation from the
EAS indicating that the physics at short distances differs from that at large distances.

In Ref.~\cite{Hue08} it is argued that in the synchrotron-emission model REAS2, the
lateral distribution function (LDF) can be used to disentangle the chemical composition of
cosmic rays. As shown in Ref.~\cite{Gou09} the predictions of the macroscopic model and the
synchrotron-emission models such as used in Refs.~\cite{Hue08,Arena52,Arena64} differ
greatly. In particular the latter models predict a unipolar pulse which contradicts the
fact that the intensity of the emitted radiation should vanish at the longest
wavelength~\cite{Arena}, a condition that is satisfied by the bi-polar pulse in the MGMR
model. Recently~\cite{Arena31,Arena34}, it has been shown that this is due to the omission
of the bremsstrahlung contributions at the beginning and the end of the particle
trajectories in the synchrotron-emission models. This effect is now included in an updated version of the geo-synchrotron code REAS3~\cite{Arena31}. Therefore, we will investigate here, as an
application of the new calculation scheme for the MGMR model, the differences in the LDF
for iron- and proton-induced showers.

In \secref{model} a quick overview of the MGMR model is given. For completeness we will
shortly review the derivation of the expression for the electric field. As the next step we
concentrate on the short distance scales in \secref{Shrt-impact} where a new expression is
derived for the electric field at small impact parameters. For simplicity, we will limit
ourselves to vertical incoming air showers. The geometry considered is an incoming shower
with velocity $\vec{\beta}=-\beta \hat{z}$, an observer placed at a distance
$\vec{d}=d\hat{x}$ from the shower axis with a magnetic field $\vec{B}=B\hat{y}$
perpendicular to the air shower. The realistic case will be treated in a forthcoming
publication. As a first application of the new calculation scheme it is shown in
\secref{angle} that the LDF depends on the orientation of the observer with respect to the
shower axis. This angular dependence is due to interference of the leading magnetic
contribution with secondary contributions. Of these secondary contributions, the one which
is generated by charge excess in the shower is the most important. As a second
application the influence of the chemical composition of the cosmic ray on the LDF is
discussed in Sections 5 and 6.

\section{The model\seclab{model}}

When an ultra-high energy cosmic ray collides in the atmosphere, a cascade of
secondary particles is created, moving towards the Earth with a velocity close to that of
the speed of light. This can be visualized by a `pancake' of particles. The basic picture
in the MGMR model~\cite{Sch08} is that the charged particles in the pancake, mostly
electrons and positrons, will be deflected in the Earth's magnetic field causing the flow
of a macroscopic electric current. The strength of this electric current is time dependent
and thus radiates. We will temporarily make the assumption that there is an equal amount of
electrons and positrons in the shower; in a later section the effects of charge excess will
be discussed.

The total amount of electrons and positrons traveling in the shower front can be described
as a function of height $z=-c\beta_s t'+h$, where $t'$ is the shower time (negative) and
the front of the shower hits the Earth at $t'=0$. This implies that the front of the shower
is located at a height of $z=-c\beta_s t'$ with the particles lagging behind at a distance
within the pancake following a distribution $f_p(h)$. The distribution of particles is
given as
\beq
 N(z,t')=  N_e f_t(t')\,f_p(h)
\eeq
where $f_t(t')$ is the normalized shower profile, and the total number of particles at the shower maximum equals $N_e=6\times(E_p/10^{10}
$eV$)$ defined by the maximum number of particles for a $10^{19}$~eV shower~\cite{Kna03}.

The longitudinal profile $N_e f_t(t')$ is parametrized~\cite{Hue03} as a function of the
penetration depth $X$ in units of g~cm$^{-2}$,
\beq
N_e f_t(t)=N_ee^{(X-X_{max}-1.5X\ln s)/X_0} \;,
\eqlab{N_e}
\eeq
The penetration depth is written as a function of height as $X(z) =
(\rho(0)/C)e^{-Cz_s}$. Using $\rho(0)\approx1168$~g~m$^{-3}$, and
$X(0)\approx1000$~g~cm$^{-2}$ gives $C=1.168\cdot 10^{-4}$~m$^{-1}$. The parameter $X_{max}$ is
taken to reproduce the shower maximum from simulations~\cite{Kna03}, 
$X_{max} = $ $(840+70\log_{10}(E_p/10^{20}\;\mbox{eV}))$ \newline$\;$ g~cm$^{-2}$. The shower age $s(X)$ can be
parameterized as~\cite{Hue03},
\beq
s(X)=\frac{3X/X_0}{X/X_0+2X_{max}/X_{0}},
\eeq
where $X_0=36.7$~g~cm$^{-2}$ is the radiation length of electrons in air.

For the pancake thickness a parametrization can be made using measured arrival time
distributions. This can be fitted using a $\Gamma$-probability distribution function
($\Gamma$-pdf)~\cite{Hue03,Agn97},
\beq
f_p(h)=h^\beta e^{-2h/L}\times(4/L^2),
\eeq
where $\beta=1$ and $L=3.9$\,m have been used, see \secref{hybrid}.

The electrons and positrons will be deflected in the Earth's magnetic field which
macroscopically induces a net current in the $\hat{x}$ direction given by,
\beq
 j^{x}(x,y,z,t')= \vd \, e\, N(z,t') \delta(x) \delta(y)\;,
\eqlab{CurrDens}
\eeq
where $q$ is the sign of the electric charge $e$. We disregard the lateral distribution of
the charged particles and fix the position of the charged particles at the shower axis. The
drift velocity, $\vd$, depends rather strongly on the model assumptions made~\cite{Wer08}
and we adopt a value of $\vd=0.025$~$c$.

The vector potential is given by the Li\'{e}nard-Wiechert fields~\cite{Jac-CE} in
terms of the current density, $j^{x}$,
\begin{equation}
 A^x(t,\vec{d})= J_0 \int
 {f_t(t_r)f(h)\over {\cal D} } \,dh\;,
 \eqlab{LW}
\end{equation}
where $J_0= \vd N_e e/\left[ 4 \pi \varepsilon_0 c \right]$, $d$ is the distance from the observer to
the shower axis, and where the retarded distance can be rewritten as
\beq
\eqlab{ret-dist}
{\cal D} = \sqrt{(-c\beta_s t+h)^2+(1-n^2\beta_s^2)d^2}  \;.
\eeq

The retarded time is defined by $c(t-t_r)/n=R$, where $R=\sqrt{(z^2+d^2)}$, $n$ is  the
index of refraction of air and can be expressed as~\cite{Sch08,Wer08},
\beq
ct_r=\frac{ct-n^2\beta_s h-n{\cal D}} {(1-n^2\beta_s^2)} \;.\eqlab{ret}
\eeq

At a large distance from the shower axis the point-like
approximation~\cite{Sch08,Gou09} is valid in which the pancake thickness is set to zero,
$f_p(h)=\delta(h)$. In this limit a simple analytical expression for the electric field can
be derived~\cite{Sch08},
\beq E_x(t,\vec{d})
 \approx J_0 {c^2 t_r^2 4 \over d^4} {d\over dt_r}[ t_r f_t(t_r)] \;,
\eqlab{E-appx}
\eeq
with $ct_r \approx -d^2 / 2 c t$. In general this approximation is valid at typical
distances above $d=500$\,m since the typical time window for the pulse lies between
$10^{-8}$ and $10^{-7}$~s. From \eqref{E-appx} we see that the electric field scales
as $d^{-4}$ and that the integral over time vanishes (i.e.\ zero response at zero
frequency) when the shower profile vanishes at the Earth's surface.

\section{Radial dependence of the pulse strength}\seclab{Shrt-impact}
For the shape of the pulse it is very important to
understand the interplay between the different length scales~\cite{Arena}. While at large
distances from the shower the important length scale is related to a projection of the
shower profile along the line of sight of the observer, we will show in this section that
for small distances the pancake thickness parameter, $L$, is important. To describe the
electric field for small impact parameters we encounter the problem that the integral in
\eqref{LW} becomes numerically unstable. We therefore first derive a new equation for the
pulse which is particularly suited at small distances from the core.

For an observer placed close to the shower axis it is more transparent to change
variables in \eqref{LW} and integrate over the retarded shower time at which the
signal is emitted since the signal moves with the same velocity as the shower.
The vector potential can thus be written as,
\bea
	A^x(t,\vec{d}) &&
 = J_0 \int\limits_{t_r^-(t)}^{t_r^+(t)} dt_r \frac{\partial h}{\partial (ct_r)} \frac{f_t(t_r)}{{\cal D}} f_p(h)\nonumber \\
  &&=-J_0 \int\limits_{t_r^-(t)}^{t_r^+(t)} dt_r \frac{f_t(t_r)}{z}f_p(h) \;. \eqlab{Ax-Shrt}
\eea
The retarded time $t_r$ is related to the observing time $t$ and the distance
from the shower front $h$ by \eqref{ret} which can be rewritten in the limit
$n=1$ and $\beta_s=1$ as
\beq
ct_r={c^2t^2-d^2-h^2 \over 2(ct+h)} \; ,
\eqlab{rettime}
\eeq
The Jacobian in this case is non-zero due to the fact that at a fixed observer time $t$, the received signal
is emitted at different shower times, depending on the emission point within the
pancake. Hence, $t_r^-=-\infty$ and $t_r^+=t_r(t,h=0)$ for a fixed observer time
$t$. The Jacobian can now be derived from~\eqref{rettime},
\beq
\frac{\partial h}{\partial (ct_r)}=\frac{{\cal D}}{ct_r-h}
\eeq
The electric field now becomes 
\bea
	E^{x}(t,\vec{d})&=&-J_0 \int\limits_{t_r^-(t)}^{t_r^+(t)}dt_r f_t(t_r) \frac{d}{dt}\left(\frac{f_p(h)}{z}\right)\nonumber\\
	&=& -J_0 \int\limits_{t_r^-(t)}^{t_r^+(t)}dt_r f_t(t_r)\left(\frac{d}{dt}+\frac{d}{dt_r}\right)\left(\frac{f_p(h)}{z}\right)\nonumber\\
	&-& J_0 \int\limits_{t_r^-(t)}^{t_r^+(t)}dt_r \frac{df_t(t_r)}{dt_r}\frac{f_p(h)}{z}\nonumber\\
	&-& \frac{f_t(t_r)f_p(h)}{z}\biggr{|}_{t_r^-(t)}^{t_r^+(t)}
\eea
Now using
\beq
\left(\frac{d}{dt}+\frac{d}{dt_r}\right)\left(\frac{f_p(h)}{z}\right) = \frac{\beta}{z}\frac{df_p(h)}{dh}
\eeq
we can write
\bea
E^{x}(t,\vec{d}) &=& -J_0 \int\limits_{t_r^-(t)}^{t_r^+(t)}dt_r f_t(t_r)\frac{\beta}{z}\frac{df_p(h)}{dh}\nonumber\\
	&-& J_0 \int\limits_{t_r^-(t)}^{t_r^+(t)}dt_r \frac{df_t(t_r)}{dt_r}\frac{f_p(h)}{z}
\eqlab{E-small}
\eea
where the terms coming from the derivative working on the integration limits and the
partial integration vanish due to the fact that the $f_p$ vanishes at the limits, and where
$t$, $t_r$ and $h$ are related by \eqref{ret}. For very small impact parameters the second
term on the right-hand side in \eqref{E-small} becomes an integral over the time
derivative of the shower profile and vanishes at the given limits. This happens since the
signal travels along with the shower and thus $h$ becomes approximately independent of
$t_r$. Hence the main contribution to the electric field for very small impact parameters
comes from the first term on the right-hand side of equation \eqref{E-small}.

\begin{figure}[!ht]
\centerline{
\includegraphics[width=.48\textwidth, keepaspectratio]{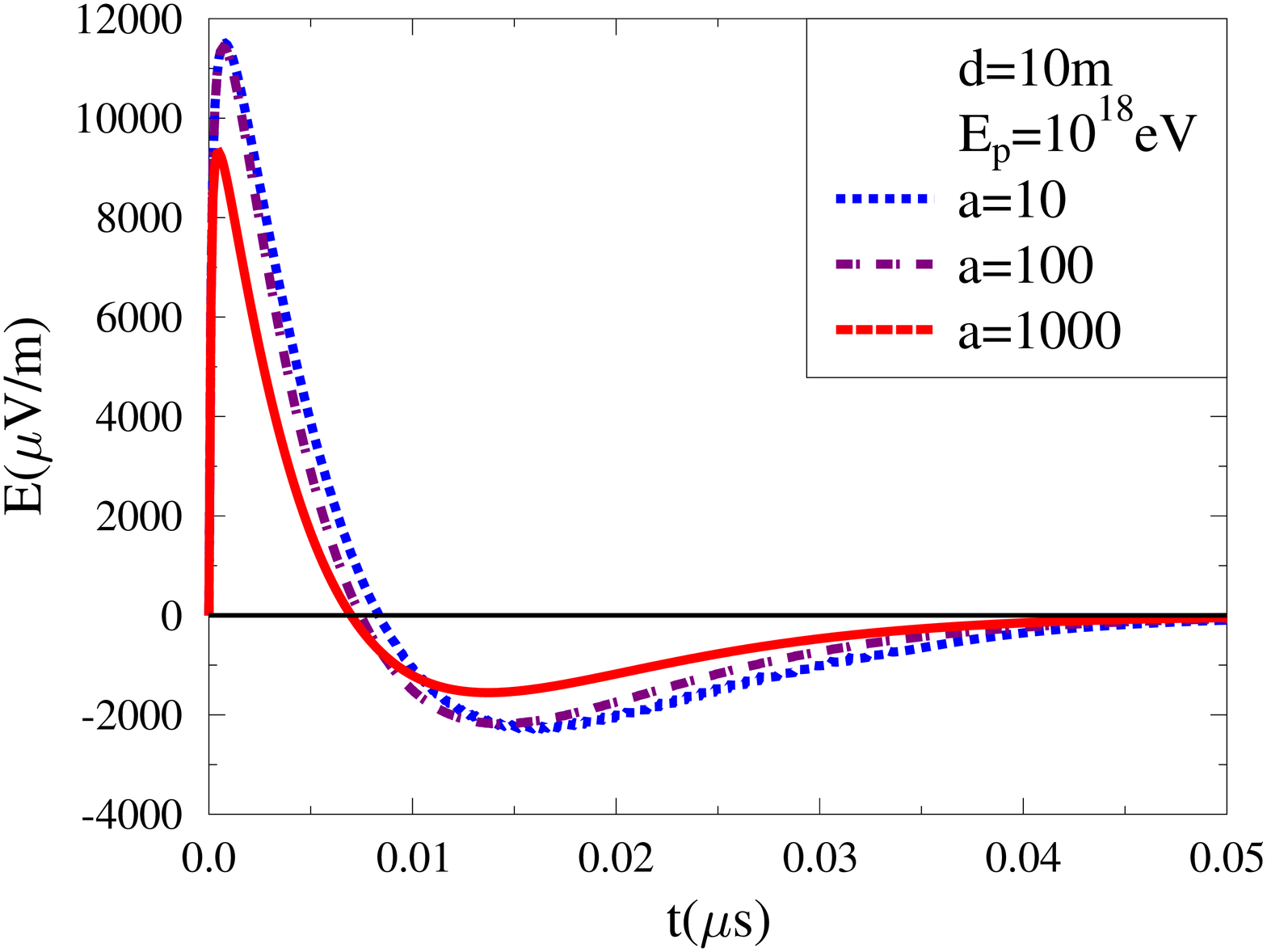}}
\centerline{
\includegraphics[width=.48\textwidth, keepaspectratio]{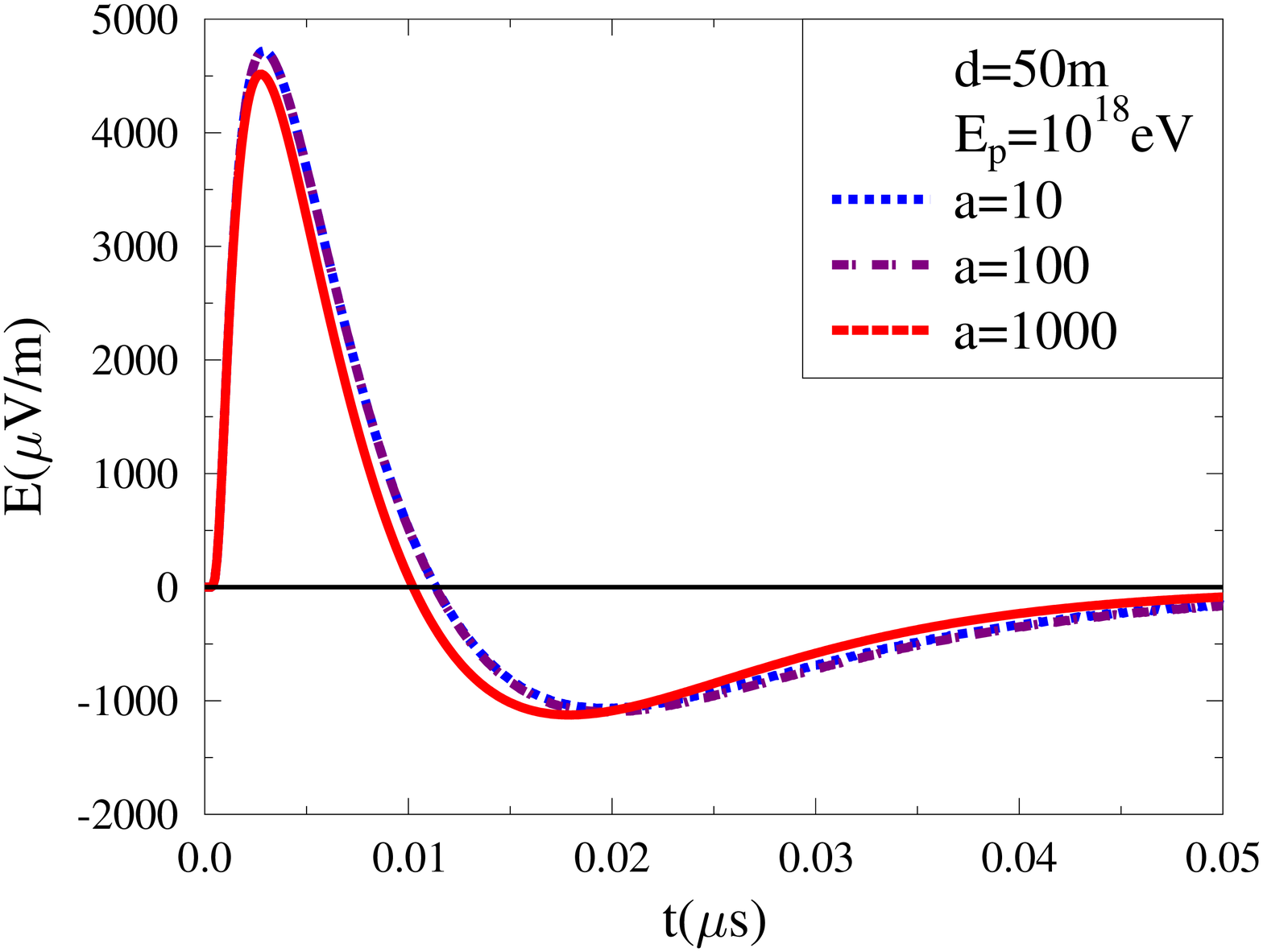}}
\caption{The unfiltered electric field for different values of the cut-off parameter $a=10$, $a=100$, and $a=1000$,
at a distance $d=10$\,m (top) and $d=50$\,m (bottom) from the shower axis.
The pancake thickness parameter is taken as $L=3.9$; see \secref{hybrid}. }
\figlab{Pulse-fall-off}
\end{figure}

When the shower profile is finite at the surface of the Earth, $z=0$, the terms in
\eqref{E-small} suffer from a $1/z$ divergence. To regularize this divergence we introduce
an exponential suppression term in the shower profile function,
\beq
F_t(t_r)=f_t(t_r)-f_t(0)e^{-z/a},
\eeq
where $a$ is the cut-off distance. The divergence in \eqref{E-small} is now eliminated
since the shower profile function vanishes linearly for $z=0$. The electric field for
different values of $a$ is given in \figref{Pulse-fall-off} where the simulations are done
for a vertical incoming shower with a primary energy of $E_p=10^{18}$\,eV. The calculations
show that at a distance of $d=50$\,m even a value as large as $a=1000$\,m hardly affects
the pulse. At a distance of only $d=10$\,m a value of $a=100$\,m gives already a stable
result. It has been verified that this result hardly depends on the altitude of the
observer or energy of the cosmic ray, and can also be used for more energetic showers where the
shower maximum is close to the Earth's surface. This is due to the fact that the maximum of
the observed electric field is emitted well before the shower is fully developed. We note here that, therefore the observation of radio signals from air showers are particularly suited to study the early stages of the shower developement. Interesting to notice is that the signal at a
distance of only 50\,m from the core is hardly affected by the shower properties at a
height below 1\,km.

Another interesting aspect of \figref{Pulse-fall-off} is that the pulse-shape does not vary
strongly with distance. At large distances where \eqref{E-appx} applies one expects a
variation like $d^2$ in pulse length. Instead one finds that the zero-crossing of the pulse
changes from 9 to 11\,ns while the distance changes by a factor 5 from $d=10$ to 50\,m.
This time scale corresponds to a length scale of about 3\,m which is determined by the
pancake thickness as supported by test calculations.
\begin{figure} 
\centerline{
\includegraphics[width=.48\textwidth, keepaspectratio]{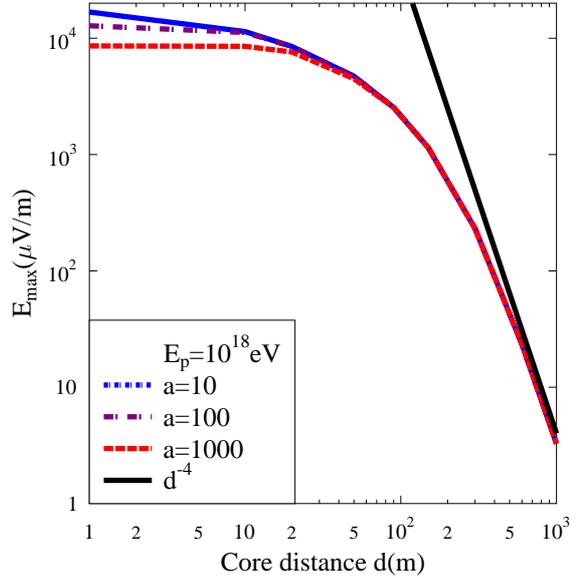}
}
\caption{Lateral distribution function for the pulse height
as function of distance on double logarithmic scale for different values of the cut-off parameter $a$.}
\figlab{Field-func-d}
\end{figure}

The dependence of the calculated pulse height on $d$ is shown in \figref{Field-func-d}.
Based on \eqref{E-appx} one would expect the pulse strength to be proportional to $d^{-4}$
as given by the black line in the figure. Only at very large distances this dependence is
seen.

\begin{figure}[!ht]
\centerline{
\includegraphics[width=.48\textwidth, keepaspectratio]{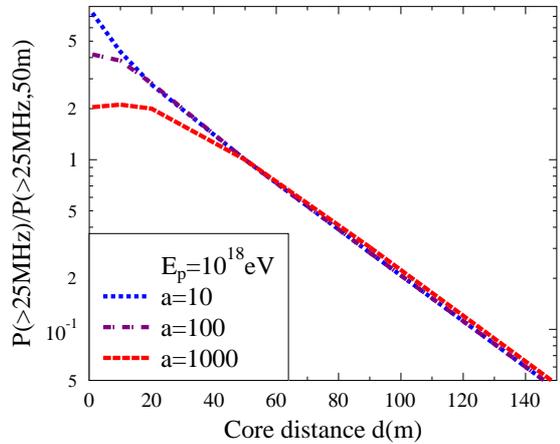}
}
\caption{Lateral distribution function for the received power at frequencies above 25 MHz divided by that
at a distance of $d = 50$~m.}
\figlab{Lat-dist}
\end{figure}

In~\figref{Lat-dist} the power $P$ above $25$\,MHz divided by that at a distance of $d =
50$~m is plotted as a function of $d$. This quantity can easily be compared to actual
measurements since the power is not affected by dispersion of the signal contrary to the
pulse height and it corresponds to a semi-realistic filtering of the pulse. It is clearly
seen that for small distances a smaller value of $a$ should be used to reach a convergent
result. The LDF for the power shows a distinct exponential structure for distances beyond
5\,m from the core where the power drops by one order of magnitude over a distance of close
to 70\,m.

For observer distances approaching $d=0$~m the pulse height tends to diverge. Since, for
$n=1$, the signal travels along with the shower with the same speed, an observer placed at
the impact point of the shower will see the equivalent of a divergent Cherenkov peak. The
precise structure of the pulse at very short distances will thus strongly depend on the
index of refraction~\cite{Wer08}. In addition one expects that at close proximity to the
core the electric field is influenced by the lateral extent of the electrons in the shower
since the particle density is strongly peaked near the shower axis. In a forthcoming
publication we will look into this effect in more detail.

\section{Interference}\seclab{angle}

The main secondary contribution, which is important for the present discussion, is due to
the charge excess in the shower~\cite{Sch08,Wer08} which is due to knock out from ambient
air molecules. The coherent emission from the charge excess in cosmic-ray induced showers
was first discussed by Askaryan~\cite{Ask62}, where it is due to the creation and annihilation of energetic electrons. Even though the Askaryan effect is often
associated with Cherenkov radiation, it also applies to the case in which the index of refraction is close to unity~\cite{Afa99}.

Macroscopically the charge-excess contribution can be described using a
net negative charge moving with the speed of light toward the surface of the Earth. The
contribution to the vector potential is now given by,
\begin{eqnarray}
A^0_{Cx}&=& \frac{eN_e}{4\pi\epsilon_0}C_x\int\limits^{\infty}_{0}dh\frac{f_t(t_r)f(h)}{{\cal D}} \nonumber \\
&-&\frac{e}{4\pi\epsilon_0}\int\limits_{-\infty}^{t'}\left( \int\limits^{\infty}_{0}dh\frac{df_t(t')}{dt'} \frac{f(h)}{R}\right) dt'\\
A^{z}_{Cx}&=&\frac{eN_e}{4\pi\epsilon_0}C_x\int\limits^{\infty}_{0}dh\frac{-\beta f_t(t_r)f(h)}{{\cal D}}.
\end{eqnarray}
The electric field is now obtained by,
\begin{eqnarray}
E_x^{Cx}(t) &=&\frac{-\partial A^0_{Cx}}{\partial x} \nonumber\\
&=&\frac{C_xeN_e}{4\pi\epsilon_0}\int\limits^{\infty}_{0}dh\frac{x}{{\cal{D}}^2}\frac{z}{R}\dot{f}_t(t_r)f(h)\\
E_y^{Cy}(t) &=&\frac{-\partial A^0_{Cx}}{\partial y} \nonumber\\
&=&\frac{C_xeN_e}{4\pi\epsilon_0}\int\limits^{\infty}_{0}dh\frac{y}{{\cal{D}}^2}\frac{z}{R}\dot{f}_t(t_r)f(h)\\
E_z^{Cx}(t) &=&\frac{-\partial A^0_{Cx}}{\partial z}-\frac{\partial A^z_{Cx}}{\partial ct} \nonumber\\
&& =\frac{C_xeN_e}{4\pi\epsilon_0}\int\limits^{\infty}_{0}dh\frac{d}{{\cal{D}}^2}\frac{d}{R}\dot{f}_t(t_r)f(h)
\end{eqnarray}
where $C_x=0.23$ is the approximate fraction of charge excess with respect to the total
number of electrons and positrons in the shower as predicted by the Monte-Carlo simulations
described in \secref{hybrid}.

\begin{figure}[!ht]
\centerline{
\includegraphics[width=.48\textwidth, keepaspectratio]{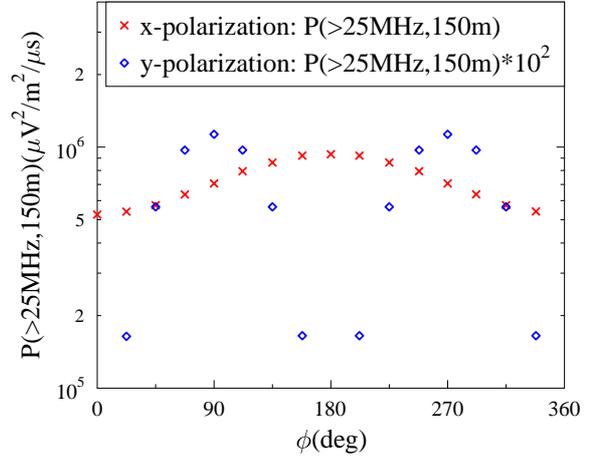}
}
\caption{The strength of the signal for two different polarization directions is plotted
as a function of the orientation of the observer with respect to the shower.
The y-polarization has been multiplied by a factor 100 to put it on a similar scale.
The magnetic field is chosen pointing to the North, while
$\phi=90^\circ$ ($\phi=180^\circ$) corresponds to the observer at the East (North) side of the shower.}
\figlab{sh-ang}
\end{figure}

The interference of the geo-magnetic and the charge-excess contribution will be important
for the observed pulse. For this interference the polarization of the two contributions
should be considered. For a vertical shower ($\hat{z}$-direction) in a horizontal magnetic
field ($\hat{y}$-direction) the electric pulse generated by the induced electric current is
polarized in the $\hat{x}$-direction independent of the orientation of the observer with
respect to the shower axis. The radiation due to charge excess is polarized radially.
Depending on the position of the observer the two contributions may interfere destructively
(observer on the negative $x$-axis, $y=0$), constructively (observer on the positive
$x$-axis, $y=0$), or be orthogonal (observer anywhere on the $y$-axis). This effect can be seen clearly in \figref{sh-ang} where the dependence of the signal strength in the $\hat{x}$-, and $\hat{y}$-direction is plotted as function of the orientation of the
observer.

\begin{figure}[!ht]
\centerline{
\includegraphics[width=.48\textwidth, keepaspectratio]{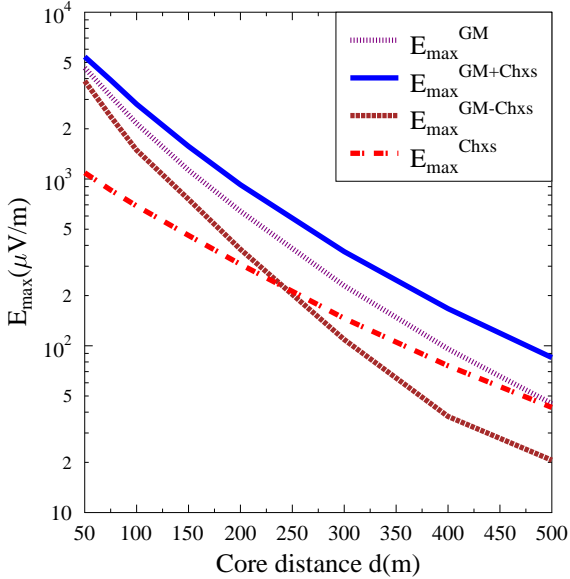}
}
\caption{The behaviour of the LDF as a consequence of different types of interference between geo-magnetic radiation and charge excess radiation.}
\figlab{ldf_int}
\end{figure}

In \figref{ldf_int} the LDF is shown separately for geo-magnetic and charge-excess radiation
showing that while the two contributions differ by a factor 5 near the shower core, their
magnitudes are about equal at a distance of 500\,m. Depending on the orientation (azimuth)
of the observer with respect to the shower the two contributions will interfere
constructively or destructively. Since the charge-excess and the geo-magnetic contribution
depend in a different way on the shower profile their LDF's differ. This gives rise to the
interesting observation that for an observer at one side of the shower the charge-excess
and the geo-magnetic contributions may interfere destructively resulting in a LDF that falls
off rather steeply close to the core and flattens after the point of maximal destructive
interference where even a local minimum could occur. For an observer positioned at the
other side of the same shower the two contributions add constructively resulting in a
relatively smooth LDF. In the perpendicular direction the two contributions will not
interfere as their polarizations are orthogonal. The distance where a local minimum could
occur and the LDF starts to flatten will depend on the relative magnitude of the charge-excess and the geo-magnetic contribution where the latter depends strongly on the angle of
the shower with the magnetic field. It follows that one should be extremely careful in
determining the LDF~\cite{Lop10} since it is expected to depend on the orientation of the
observer to the shower.

\section{Hybrid approach}\seclab{hybrid}

In the previous sections, all simulations were done using the basic MGMR model. The particle distributions in these simulations were parametrized using the analytical formula given in~\secref{model}. In the current section we will discuss the hybrid approach to the MGMR model. Within this approach the important macroscopic properties describing the electric field, such as the pancake thickness and the shower profile, are obtained from Monte-Carlo simulations.

These simulations are done
using the cascade mode of the CONEX shower simulation
program~\cite{CON1,CON2}. This code has been modified to include
the deviation of charged particles in the Earth's magnetic
field, which is discussed in detail in Appendix B of
Ref.~\cite{Wer08}. Furthermore an analysis tool has been written, giving
a histogrammed output of the full three-dimensional and
timing information of the currents and particle distributions
within a user-defined region of the shower front which
in the simplest case becomes an average of all particles
over the complete shower front. Several examples for such
distributions, smoothed by fitting procedures, can be found in
Ref.~\cite{Wer08}, more details will be discussed in a future publication.
The complete shower-simulation package including analysis tools
is called CONEX-MC-GEO.
\begin{figure}[hhc!]
\centerline{
\includegraphics[width=.48\textwidth,keepaspectratio]{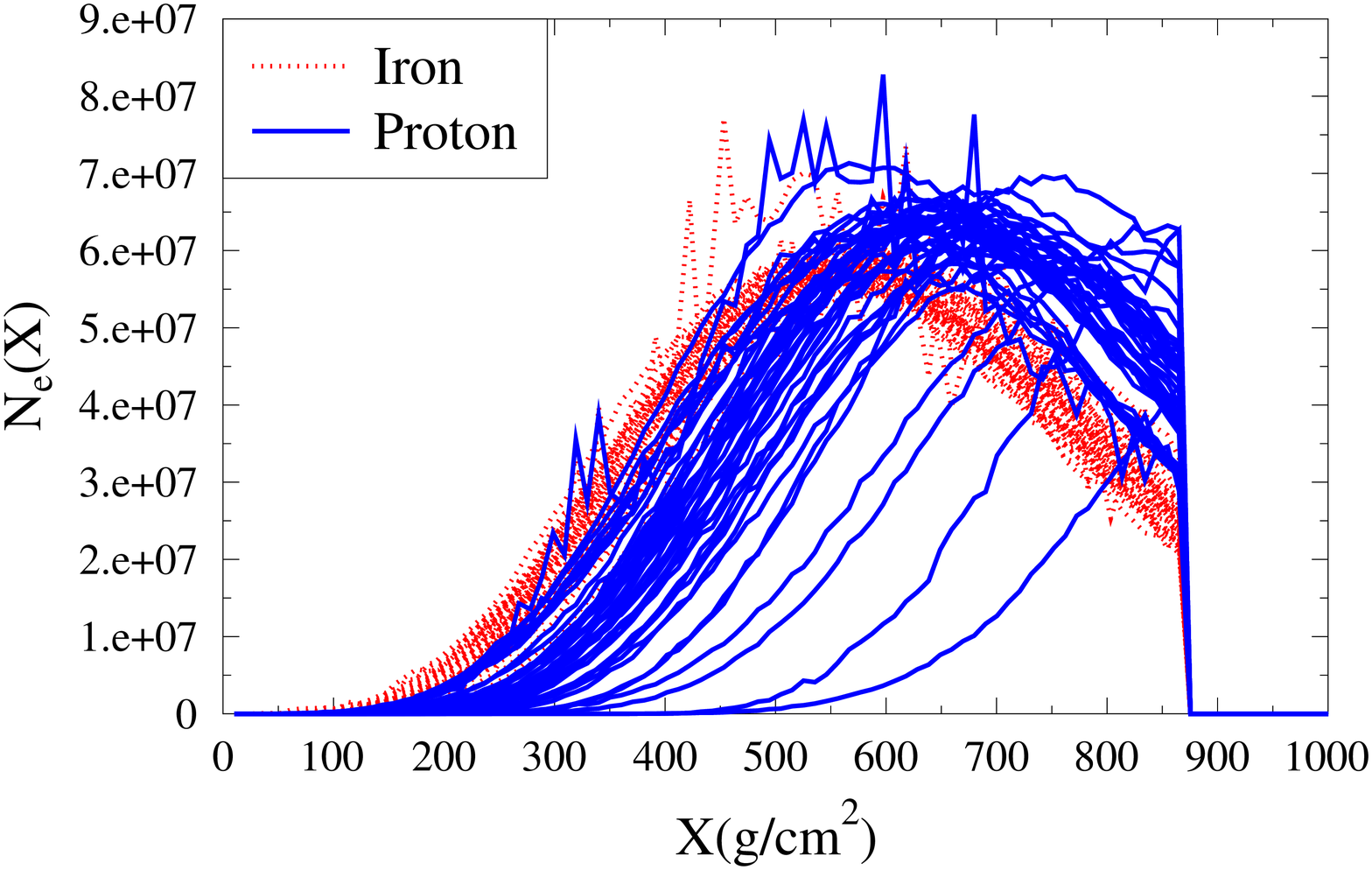}}
\centerline{
\includegraphics[width=.48\textwidth,keepaspectratio]{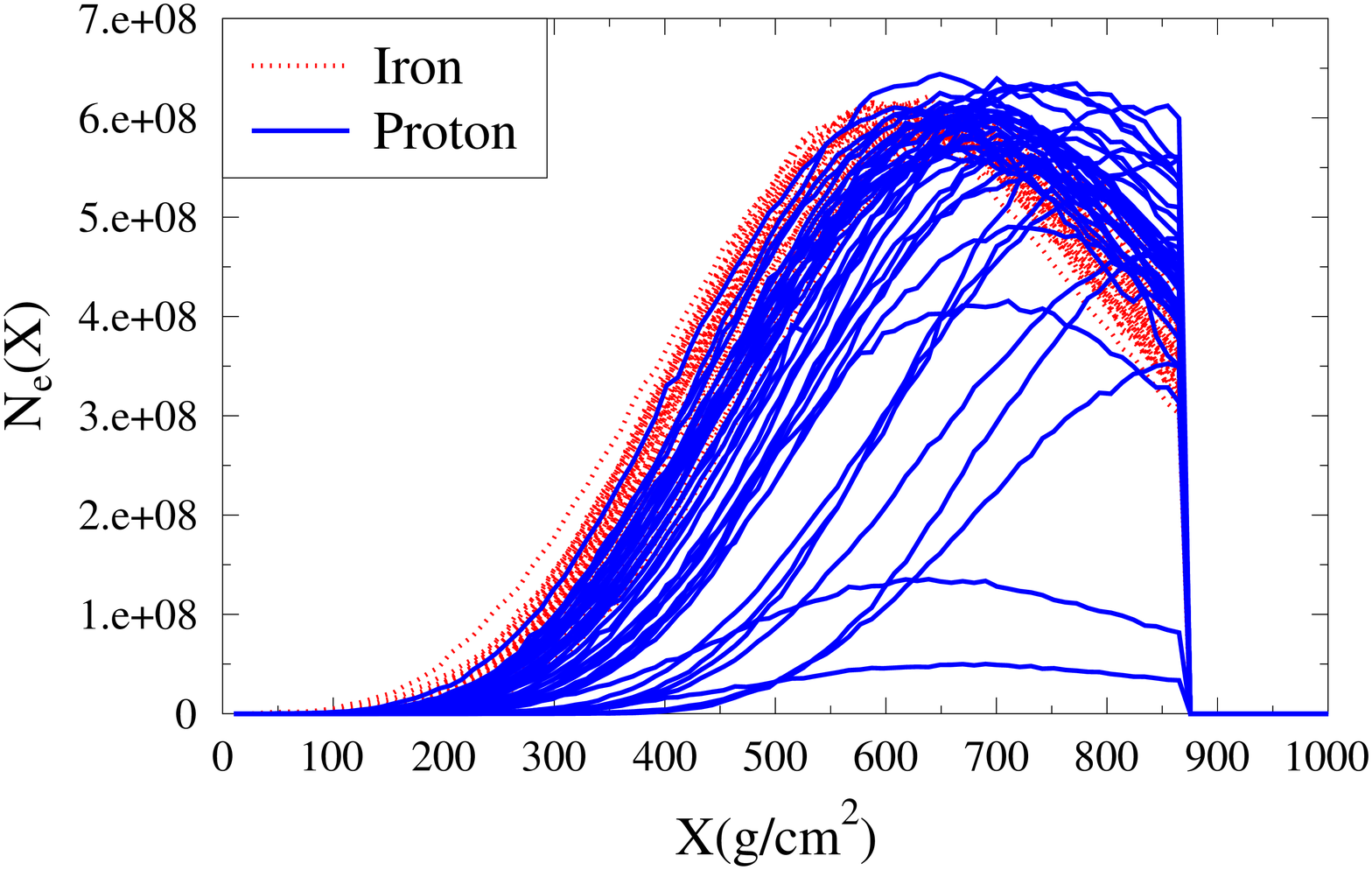}
}
\caption{Total number of electrons and positrons as function of shower depth
for 40 proton (blue curves) and 40 iron (red curves) initiated showers at energies
$E=10^{17}$\,eV (top) and $E=10^{18}$\,eV (bottom). }
\figlab{sh-prof}
\end{figure}

As a first example the histogrammed output obtained by CONEX-MC-GEO, which will be used in~\secref{mass} to study the differences of the LDF for a proton- and an iron-primary particle, will be discussed. The simulations are done for 40 proton- and 40 iron-induced
showers to extract the longitudinal shower profiles and the pancake thicknesses at energies of $10^{17}$ and $10^{18}$\,eV. 

The obtained shower profiles are shown in~\figref{sh-prof}.
As is expected the shower-to-shower fluctuations are rather large for proton- and small for
iron-induced showers. It should be noted that the intrinsic fluctuations for a few showers become large. This is partially due to thinning effects. The $X_{max}$ is larger for proton than for iron. The shower profile
drops to zero at around $870$\,g~cm$^{-2}$ where the shower hits the surface of the Earth
according to the conditions at the site of the Pierre Auger Observatory.

In the previous section the importance of the charge-excess contribution and its impact on
the LDF is shown. To make this more quantitative the fraction of excess electrons is 
plotted in \figref{chxxs} for Monte-Carlo simulations of 40 proton- and 40 iron-induced
showers at an energy of $10^{17}$\,eV. An analysis for $10^{18}$\,eV showers gives similar
results. The charge excess is slightly increasing with increasing shower depth. Near the
shower maximum the charge excess is close to $23\%$ of the total number of electrons and
positrons which is the value we have assumed in all our calculations described in~\secref{angle}. The charge-excess contribution scales linearly with the charge-excess fraction. The shower-to-shower fluctuations thus have a relatively small effect.
\begin{figure}[hhc!]
\centerline{
\includegraphics[width=.48\textwidth,keepaspectratio]{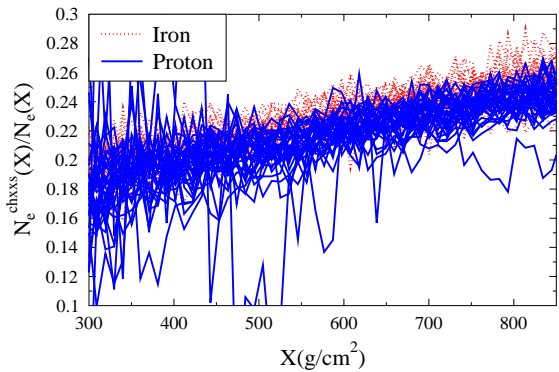}}
\caption{The fraction of charge-excess electrons with respect to the total number of electrons and positrons
as a function of shower depth for 40 proton and 40 iron induced showers at an energy of $10^{17}$\,eV.  }
\figlab{chxxs}
\end{figure}
The other important parameter for the radio-emission calculations is the pancake thickness
parameter. The relevant parameter for this is the mean distance of the electrons behind the
shower front, $\left\langle h\right\rangle$, the mean pancake thickness parameter. In
\figref{sh-pan} this quantity is plotted for the set of showers at $10^{17}$\,eV.
Again the fluctuations are considerably larger for protons than for iron. In
addition one can see that the mean pancake thickness parameter is almost independent of shower height and of shower type.
Calculations for showers for $10^{18}$\,eV give similar results.

In the present calculations $L=\left\langle h \right\rangle$ has been fixed for the full
shower development. From \figref{sh-pan} it can be seen that $L=4.3$\,m is a reasonable
average value for iron-induced air showers, and $L=3.9$\, for proton-induced air showers. Where the width of the function at $1/e$ of the maximum can be shown to be
approximately $1.5L$. The fluctuations in the average value of $L$ are small with respect to the fluctuations in the shower profiles and not considered in further analysis.
\begin{figure}[!ht]
\centerline{
\includegraphics[width=.48\textwidth, keepaspectratio]{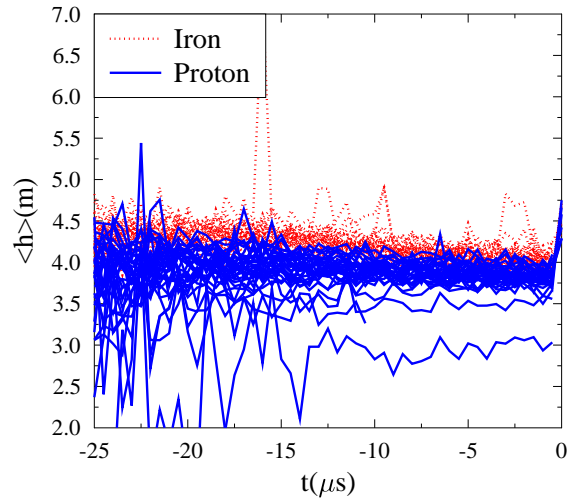}
}
\caption{Mean distances of electrons behind the shower front for iron
(red curves), and proton (blue curves) induced showers, as function of the
negative shower time.}
\figlab{sh-pan}
\end{figure}

\section{Composition}\seclab{mass}
As a first application of the hybrid approach we study the LDF for proton- and iron-induced showers using 
MGMR in combination with CONEX-MC-GEO. The smoothed shower profiles extracted from each Monte-Carlo simulation are used in the calculation of the radio signal. For the calculation at short distances we use a suppression coefficient of $a=100$, since in~\secref{Shrt-impact} it is shown that this
value gives reliable results at the distances considered here. The
maximum of the field strength as function of distance is shown in \figref{sh-ldf_FeH}. This is
done for an observer placed on the $x$-axis where the charge-excess and geo-magnetic fields
interfere constructively. This figure shows that the slope for iron-induced showers is less
steep than for protons. One also sees that at a distance of $d=100$\,m the calculated
signal strength for iron and proton showers is almost the same. These features were also
observed in REAS2 calculations~\cite{Hue08}.

\begin{figure}[!ht]
\centerline{
\includegraphics[width=.48\textwidth, keepaspectratio]{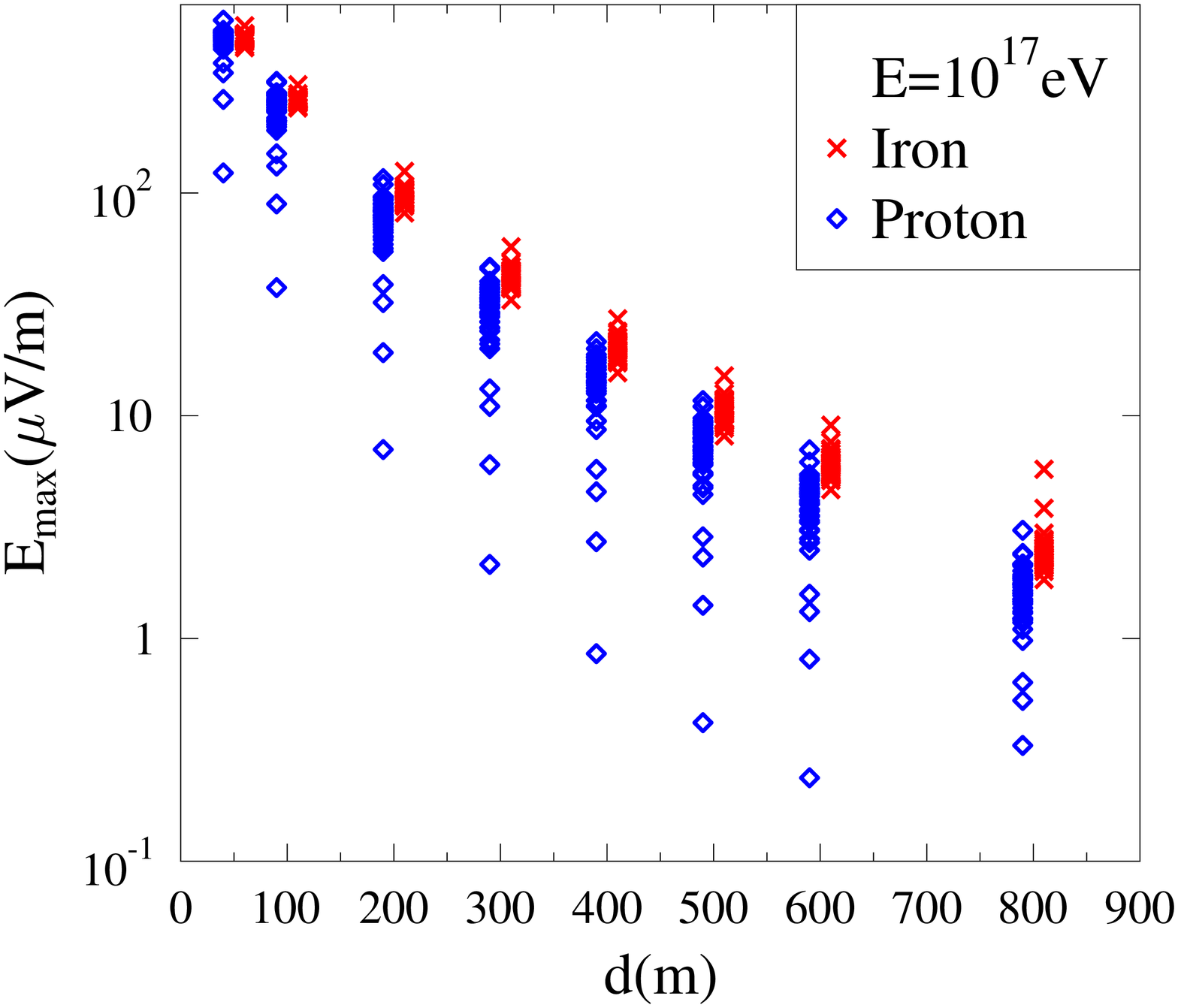}}
\centerline{
\includegraphics[width=.48\textwidth, keepaspectratio]{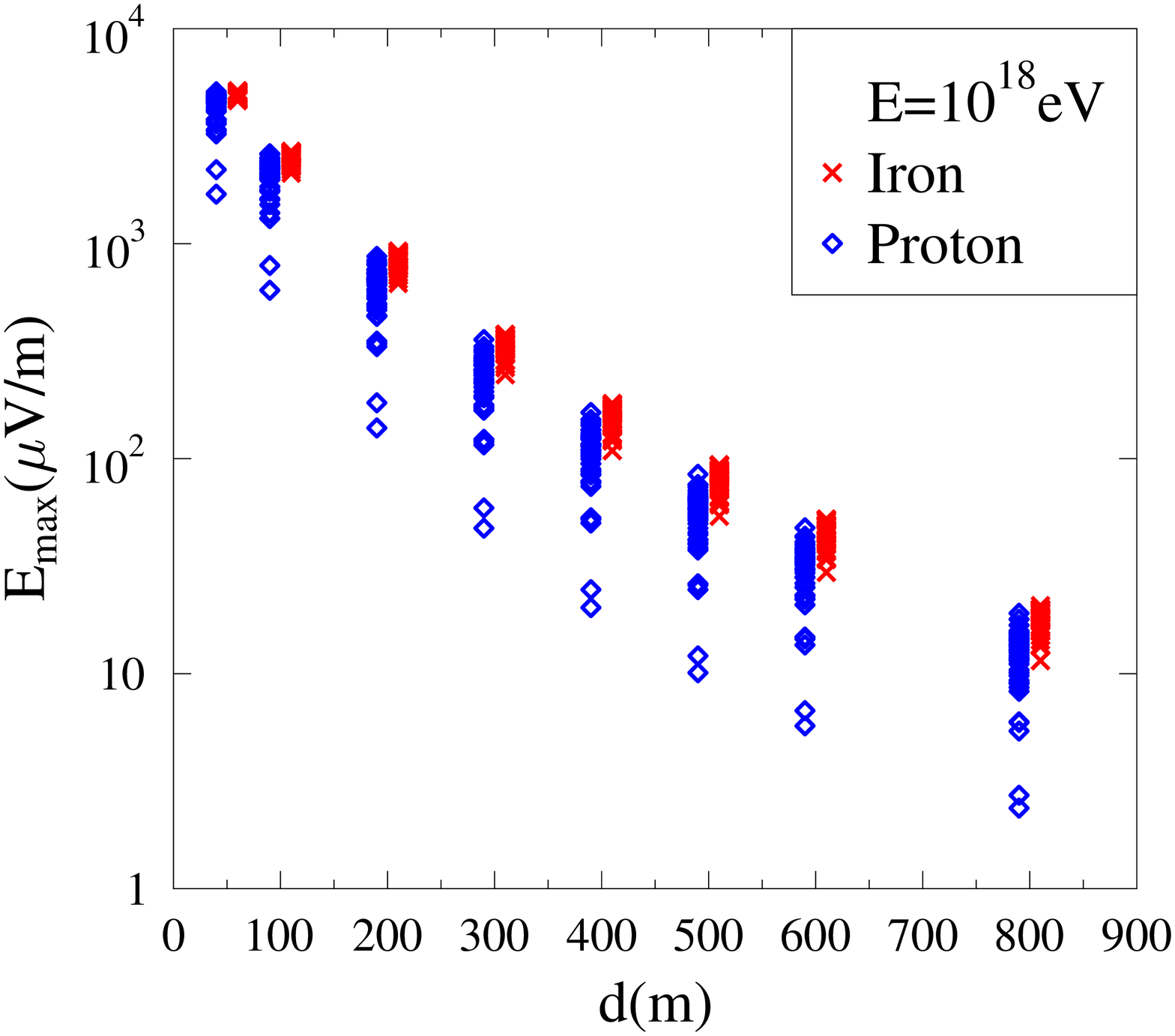}}
\caption{The LDFs are shown for proton (blue diamonds) and iron (red crosses) induced
showers at an energy of $10^{17}$\,eV (top), and $10^{18}$\,eV (bottom), displayed in \figref{sh-prof}. The LDFs are given for an observer position leading to constructive interference between the charge excess contribution and the contribution due to geo-magnetic radiation.}
\figlab{sh-ldf_FeH}
\end{figure}

To express the differences in the LDF more clearly we introduce a variable that
is relatively easy to extract from data,
\beq
R^{25}_{50/300}=P(50,f>25)/P(300,f>25)
\eeq
where $P(d,f>25)$ is the power in the pulse for frequencies larger than 25\,MHz at a
distance $d$\,m from the shower core. 
In \figref{sh-pwr} the results of
the calculations are displayed in a histogram. The averages of the values for iron- and
proton-induced showers differ strongly, however due to shower-to-shower fluctuations some
of the proton showers give rather similar values as for iron. The proton showers that give
a similar value for $R^{25}_{50/300}$  are showers which have profiles (see
\figref{sh-prof}) very similar to those of iron. At higher energies the average value of
$R^{25}_{50/300}$ differs less between proton- and iron-induced showers. This is a direct
consequence of the fact that $X_{max}$, the important parameter responsible for the
observed effect, differs less for the different primary particles at higher energies, as
can be seen in~\figref{sh-ldf_FeH}. As noted earlier the LDF, and thus also
$R^{25}_{50/300}$, depends on the orientation of the observer with respect to the shower
and comparisons should thus be made at the same azimuth with respect to the shower core. It
has been checked that the qualitative differences between iron and protons remain as shown
in \figref{sh-pwr} also for other azimuth positions of the observer. The reason for this is
that the charge-excess fraction is similar for proton- and iron-induced showers.

\begin{figure}[hhc!]
\centerline{
\includegraphics[width=.48\textwidth, keepaspectratio]{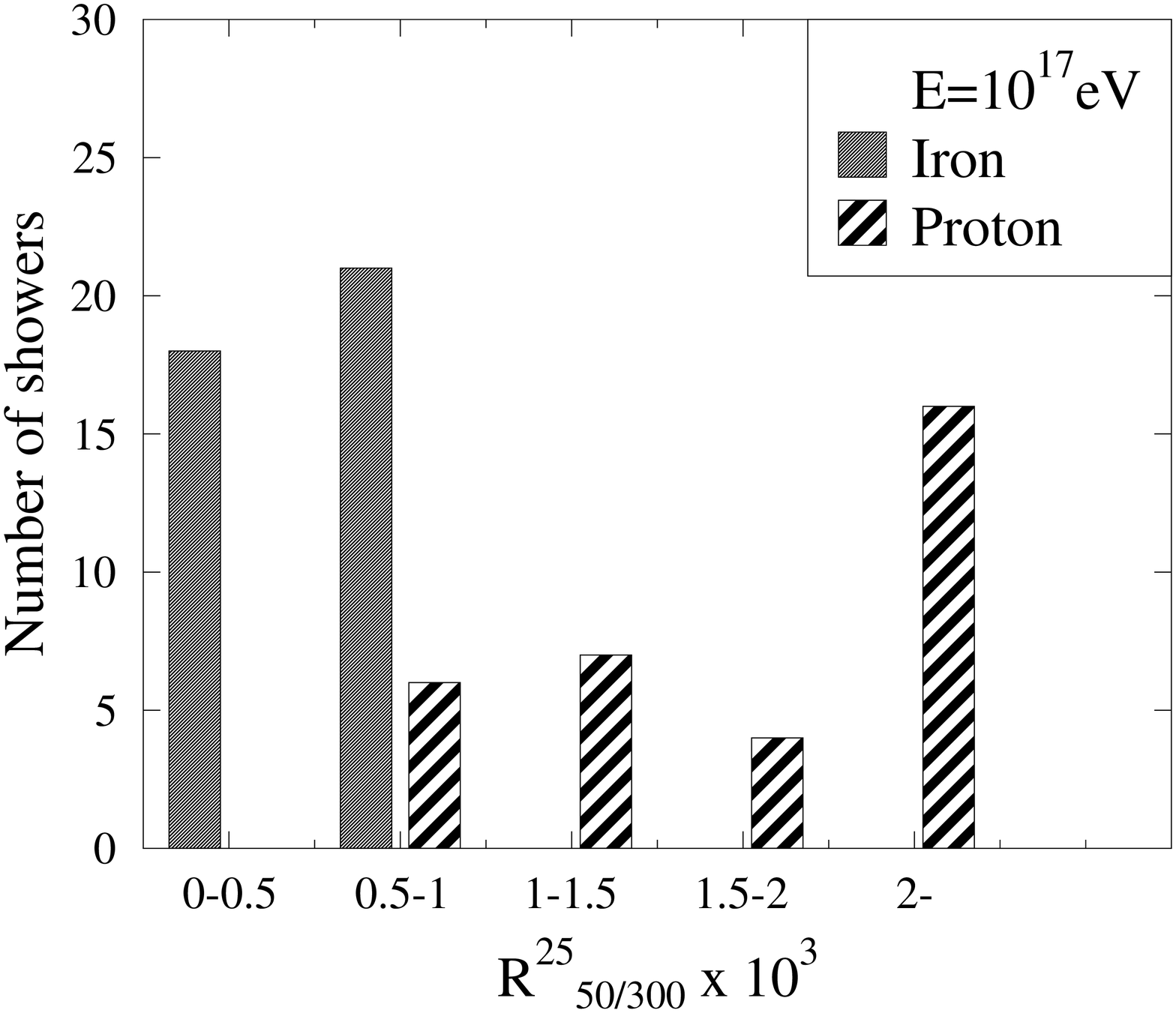}}
\centerline{
\includegraphics[width=.48\textwidth, keepaspectratio]{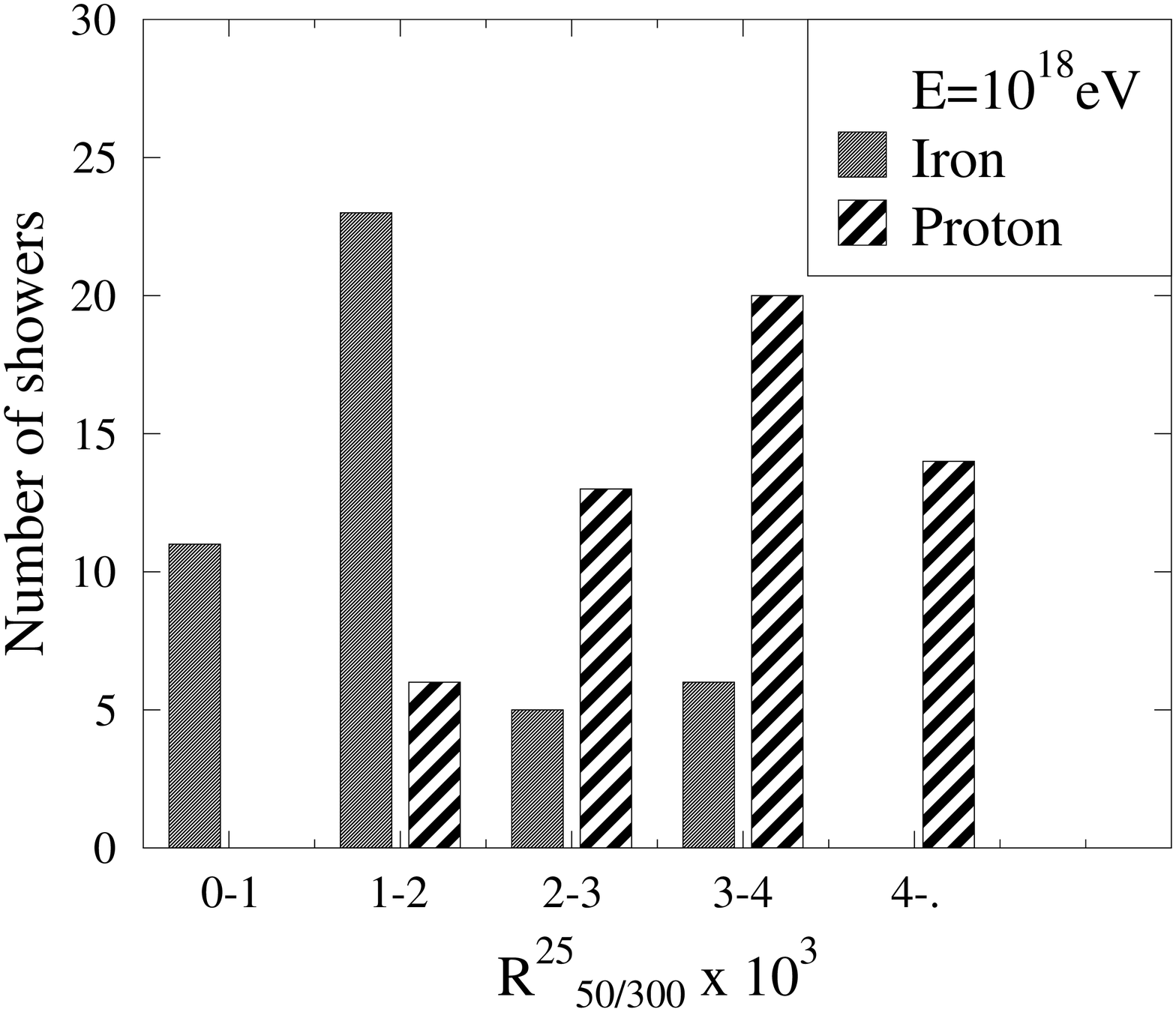}
}
\caption{$R^{25}_{50/300}$, see text, for the 40 simulated proton and iron showers at an energy of $E=10^{17}$\,eV (top) and $E=10^{18}$\,eV (bottom).}
\figlab{sh-pwr}
\end{figure}

\section{Summary and Conclusions}

We have derived a new expression for the electric field emitted by an EAS at close
proximity to the shower axis in the MGMR model. It differs from the expression used for the
pulse strength at large distances by a change of integration variables. This
change of variables expresses that at large distances the pulse shape is basically
determined by the shower profile where the pancake thickness is integrated over. At short
distances the picture is reversed and the pulse shape is determined by the pancake function
and the shower profile can be integrated.

Due to the fact that the electron density is finite when the shower reaches Earth
the pulse height becomes divergent at short distances from the core. This we have
resolved by -artificially- suppressing the shower at small heights above the
surface of the Earth. We show that this allows us to get accurate numerically-stable
results at distances down to 30\,m. At smaller distances other effects such
as the lateral spread of the electrons in the shower, as well as shower-to-shower
fluctuations will be important which are not included in the present schematic
approach. In a future publication these effects will be considered in detail.

Since at short distances, the pulse shape is determined by the pancake function while at
large distances it is governed by the shower profile this offers the possibility to use the
LDF of the radio pulse to distinguish iron- and proton-induced showers. Before addressing
this point we have added the radiation due to charge excess as the most important secondary
process. It is shown that the determination of the polarization as a function of the
azimuth angle can be used to study the different emission mechanisms of radio emission from
an EAS. We show that the LDF is a very powerful tool to distinguish iron- and proton-induced
showers although the effects of shower-to-shower fluctuations for protons can be large,
where at the same time the power at a distance of $50$ m from the shower core can be used
as a measure for the energy.

\section{Acknowledgment}
This work is part of the research program of the `Stichting voor Fundamenteel
Onderzoek der Materie (FOM)', which is financially supported by the `Nederlandse
Organisatie voor Wetenschappelijk Onderzoek (NWO)'.


\begin{thebibliography}{00}

\bibitem{Hue08}
    T. Huege, R. Ulrich, R. Engel, Astropart. Phys. \vyp{30}{2008}{96}.

\bibitem{Sch08} Olaf Scholten, Klaus Werner, and Febdian Rusydi, Astropart.\
    Phys.\ \vyp{29}{2008}{94}.

\bibitem{Wer08} Klaus Werner and Olaf Scholten, Astropart. Phys.\
    \vyp{29}{2008}{393} 

\bibitem{Fal05} H. Falcke, et al.,
Nature \vyp{435}{2005}{313}.

\bibitem{Ape06} W.D. Apel, et al.,
Astropart. Physics \vyp{26}{2006}{332}.

\bibitem{Ard06} D. Ardouin, et al.,
Astropart. Physics \vyp{26}{2006}{341}.

\bibitem{Ard09} D. Ardouin and the CODALEMA Collaboration, Astropart.Phys.,\
   \vyp{31}{2009}{192}.
\bibitem{Fal03} H.~Falcke and P.~Gorham, Astropart. Phys. \vyp{19}{2003}{477}.

\bibitem{Sup03} 
D.A. Suprun, P.W. Gorham, J.L. Rosner, Astropart. Phys. \vyp{20}{2003}{157}.

\bibitem{Hue05} T. Huege, H. Falcke, Astronomy \& Astrophysics
    \vyp{430}{2005}{779}

\bibitem{Jel65} J.V. Jelley, et al., Nature \vyp{205}{1965}{327}.

\bibitem{Por65} N.A. Porter, C.D. Long, B. McBreen, D.J.B Murnaghan and T.C.
    Weekes, \pl{19}{1965}{415}.

\bibitem{Kah66} F.D. Kahn and I. Lerche, Proc. Royal Soc. London
    \vyp{A289}{1966}{206}.

\bibitem{All71} H.R. Allan, 
Prog. in Element. part. and Cos. Ray Phys. \vyp{10}{1971}{171}.

\bibitem{Ber07} A.M. van den Berg for the Pierre Auger Collaboration, Proc. of the 30th ICRC, Merida, Mexico \vyp{5}{2008}{885}

\bibitem{Cop09} J.~Coppens, Pierre Auger Coll., Nucl.~Instr.~and Meth.~A \vyp{604} {2009} {S41}.


\bibitem{Rev09} B.~Revenu, Pierre Auger Coll., Nucl.~Instr.~and Meth.~A \vyp{604} {2009} {S37}

\bibitem{Arena} O.~Scholten and K.~Werner, Nucl.~Instr.~and Meth.~A \vyp{604} {2009} {S24}.

\bibitem{Gou09} T. Gousset, J. Lamblin, and S. Valcares, Astropart. Phys.\
    \vyp{31}{2009}{52}.

\bibitem{Arena52} C. Riviere,  Contribution nr.\ 52 to the ARENA 2010
    conference, Nantes, to be published.

\bibitem{Arena64} V. Marin,  Contribution nr.\ 64 to the ARENA 2010
    conference, Nantes, to be published.

\bibitem{Arena31} M. Ludwig, and T. Huege, Contribution nr.\ 31 to the ARENA 2010
    conference, Nantes, to be published.

\bibitem{Arena34} T. Huege, et al.,  Contribution nr.\ 34 to the ARENA 2010
    conference, Nantes, to be published.

\bibitem{Kna03} J. Knapp et al.,
Astropart. Phys. \vyp{19}{2003}{77} 

\bibitem{Hue03} T. Huege, H. Falcke, Astronomy \& Astrophysics
    \vyp{412}{2003}{19}.

\bibitem{Agn97} G. Agnetta et al.,
Astropart. Phys. \vyp{6}{2003}{301} 

\bibitem{Jac-CE} J.D. Jackson, {Classical Electrodynamics}, Wiley, New
    York, 1999.

\bibitem{Ask62} G.A. Askaryan, Sov. Phys. JETP \vyp{14}{1962}{441}; \vyp{21}{1965}{658}.

\bibitem{Afa99} G.N. Afanasiev, V.G. Kartavenko, and Yu.P. Stepanovsky, J. Phys. D \vyp {32}{1999Í’}{2029}




\bibitem{Lop10} W.D. Apel et al. -LOPES collaboration,
Astropart. Phys. \vyp{32}{2010}{294}

\bibitem{CON1}  G. Bossard, H.J. Drescher, N.N. Kalmykov, S. Ostapchenko, A.I.
  Pavlov, T. Pierog, E.A. Vishnevskaya, and K. Werner,
  Phys.Rev. \vyp{D63}{2001}{054030}


\bibitem{CON2} T. Bergmann, R. Engel, D. Heck, N.N. Kalmykov, Sergey
  Ostapchenko, T. Pierog, T. Thouw, K. Werner,
  Astropart.Phys. \vyp{26}{2007}{420} 


\end{thebibliography}
\end{document}